\documentclass[floatfix,aps,pre,showpacs]{revtex4}
\pdfoutput=1
\usepackage{graphicx}
\usepackage{amsmath}
\usepackage{amsfonts}
\usepackage{amssymb}
\usepackage{rotating}
\usepackage{fancyhdr}
\usepackage{lastpage}
\usepackage{color}

\begin{document}
\pagestyle{fancy}
\fancyhead{}
\fancyfoot{}
\fancyhead[LO,RE]{\slshape \leftmark }

\renewcommand{\headrulewidth}{0.4pt}
\renewcommand{\footrulewidth}{0.4pt}
\renewcommand{\headrulewidth}{0pt}

\newcommand{\cut}[1]{}
\newcommand{\old}{\textcolor{red}}
\newcommand{\new}{\textcolor{blue}}

\title{Generalized mean field approximation for parallel dynamics of the Ising model }

\author{Hamed Mahmoudi }
\author{David Saad }
\affiliation{ Non-linearity and Complexity Research Group, Aston University, Birmingham B4 7ET, United Kingdom }


\begin{abstract}
The dynamics of non-equilibrium Ising model with parallel updates is investigated using a generalized mean field approximation that incorporates multiple two-site correlations at any two time steps, which can be obtained recursively. The proposed method shows significant improvement in predicting local system properties compared to other mean field approximation techniques, particularly in systems with symmetric interactions. Results are also evaluated against those obtained from Monte Carlo simulations. The method is also employed to obtain parameter values for the kinetic inverse Ising modeling problem, where couplings and local fields values of a fully connected spin system are inferred from data.
\end{abstract}

\maketitle

\section{Introduction}
Statistical physics of infinite-range Ising models has been extensively studied during the last four decades~\cite{Parisi-RSB1,MezardParisiVirasoro,sompol1}. Various methods such as the replica and cavity methods have been devised to study the macroscopic properties of Ising systems with long range random interactions at equilibrium. Other mean field approximations that have been successfully used to study the infinite-range Ising model are naive mean field and the TAP approximation~\cite{MezardParisiVirasoro,KabashimaSaad}, both are also applicable to broad range of problems~\cite{Frontiers,opper2001advanced}.

While equilibrium properties of Ising systems are well-understood, analysing their dynamics is still challenging and is not completely solved. The root of the problem is the multi-time statistical dependencies between time steps manifested in the corresponding correlation and response functions.
Early attempts to analyze the dynamics of spherical spin models via mean field methods were based on the Langvin equations~\cite{sompol1,sompol2} while a path integral analysis of the Glauber dynamics in spin glass systems was developed in~\cite{sommers,biroli1999dynamical}. More recently, motivated by the need to infer parameter values in the kinetic inverse Ising modeling problem, several naive mean field approximations has been proposed for studying the dynamics of Ising system~\cite{Mezard2011exact,sakellariou2012effect,HertzRoudi2011}.

The Ising spin dynamics is usually described by a Markovian process whereby at every time step each spin $\sigma_i$ is updated according to a probability that depends on effective field $h_i=\sum_j J_{ji} \sigma_j$. For the infinite range model, the interaction couplings $J_{ij}$ are random variables sampled from a Gaussian distribution and the strength of couplings decreases with the system size. In symmetric networks {\it i.e.}, where $J_{ij}=J_{ji}$, the stationary state of system under Glauber dynamics can be described by Gibbs measure, a property that does not hold for asymmetric networks. In both cases, an \emph{exact} transient analytical solution is unfeasible, and the evolution of dynamical quantities of interest such as magnetizations and correlations can be obtained only by time consuming numerical simulations.

It is therefore essential to develop accurate approximation methods for better understanding the dynamics of spin systems as well as for applications that require probabilistic inference of microscopic variables. One approach is to start from fully asymmetric networks which exhibit small correlations among spins at different times. This allows one to use central limit theorem and apply the Gaussian approximation for probability distributions of effective fields. Based on this property, an exact formalism was introduced for the dynamics of \emph{fully asymmetric networks} in~\cite{Mezard2011exact}.

A central macroscopic quantity in studying the dynamics of spin glasses is the spin correlation function that describes the correlation between microscopic states at two time steps. In the present work we extend the mean field approximation introduced in~\cite{Mezard2011exact} by incorporating information from the time dependent correlations. We still consider the effective fields to follow a Gaussian distribution but modify the standard deviation of the effective fields by considering the non-zero covariance between spins at two different time steps. We show how this modification improves the results, particularly for symmetric networks. Moreover, we provide an analytical recursive equation for calculating the covariance among spins at different time steps, a property which has been absent from other mean field methods.

To demonstrate the applicability of our approach as an inference tool for dynamical data we apply it for studying the network reconstruction problem. The latter is an inverse problem in which one assumes to have access to dynamical data from which time-dependent quantities such as magnetization and correlations could be calculated. The challenge is to infer model variable values that best describe the sample data; the model used in the current case is the Ising model, hence microscopic interaction and external field values should be inferred.

In the original equilibrium inverse Ising problem, the assumption is that the data is drawn from
a Boltzmann-Gibbs distribution with an energy function that includes symmetric pairwise interactions~\cite{Schneidman05,Frontiers}.
In the non-equilibrium variant of the problem~\cite{Mezard2011exact,zeng2011network,HertzRoudi2011} dynamical mean field methods are used to infer the best interaction values that describe the data where samples are time-dependent.
This problem has been actively studied in recent years, in particular in the context of
neural network inference and protein-protein interactions~\cite{Roudi09,WeigtPNAS2009}.

The paper is organized as follows. In Sec.~\ref{sec:model} we briefly describe the model, introduce the macroscopic observables and the dynamic update rules. In Sec.~\ref{sec:infinite}
we use a Gaussian approximation which is exact for the fully asymmetric networks in the infinite-range model, but incorporate covariance values between spins at different time steps; we also derive a method for recursively calculating covariance values for spins at all times.
In Sec.~\ref{sec:inverse} we introduce the kinetic inverse Ising problem and the corresponding application of our method as an inference tool. Section~\ref{sec:results} focuses on the numerical results obtained for the two problems: (1) the forward calculation of macroscopic properties; and (2) inferring model values for the inverse Ising problem. We compare our results to other mean field approximation for both cases as well as against direct numerical simulation observed in Monte Carlo simulations.
Section(\ref{sec:conclusion}) introduces concluding remarks and offers future research directions.

\section{Model}
\label{sec:model}
We consider an Ising spin system comprising $N$ binary variables
$\vec{\sigma} =\{\sigma_1,..,\sigma_{N}\}$ which are connected through random couplings $J_{ij}$. The interactions can be asymmetric $(J_{ij}\neq J_{ji})$ or symmetric $(J_{ij}=J_{ji})$. Following~\cite{sompol1,sompol2} we introduce a parameter $k$ which measures the degree of symmetry in the couplings
\begin{equation}
J_{ij} = J_{ij}^{\rm{s}} + k J_{ij}^{\rm{as}}~,
\end{equation}
where $J^{\rm{s}}$ and $J^{\rm{as}}$ are symmetric and antisymmetric matrices, respectively.
The off-diagonal elements of $J_{ij}~\forall i,j$  are random
Gaussian variables with zero means and variance
\begin{equation}
\left[(J_{ij}^{\rm{s}} - J_0/N)^2\right]_{J} = \left[(J_{ij}^{\rm{as}})^2\right] = \frac{1}{N^2} \frac{1}{1+k^2}~.
\end{equation}
We define the self connectivity to be zero $(J_{ii} =0)$. The parameter $k=0$ corresponds to fully symmetric networks whereas $k=1$ represents fully asymmetric networks. The interactions among spins determine the dynamics of system; in the parallel update scheme, all spins are updated simultaneously
\begin{eqnarray}
\sigma_i(t+1) = \left\{
\begin{array}{c cl}
+1 &{\rm with \,\,probability} \,\,\,\, \frac{1}{1 + e^{-2\beta(h_i(t+1)+\theta_i(t+1))}}&\\
-1 &{\rm with \,\,probability} \,\,\,\, \frac{1}{1+e^{2\beta(h_i(t+1)+\theta_i(t+1))}}& ~,
\end{array}
\right.
\label{eq:dynamic}
\end{eqnarray}
where $h_i(t)$ is the effective local field acting on spin $i$ at time step $t$
\begin{equation}
h_i(t) = \sum_{j\in\partial i} \, J_{ji}\,\sigma_j(t-1)+ \theta_i(t) \,\,\,\,\, ,
\label{eq:eff_field}
\end{equation}
and the parameter $\beta$, analogous to inverse temperature, is
a measure of the overall strength of the interactions. The notation $\partial i$ stands for the set of spins neighboring target spin $i$ with $J_{ji}\neq 0$. In this paper we only study the fully connected networks; however, this notation is use for consistency. This particular choice of sequential update [Eq.~(\ref{eq:dynamic})] is known as {\it Glauber dynamics}.

In this paper we focus only on synchronous updates and aim to derive local system properties such as magnetizations and correlations directly from the Glauber dynamics. The joint probability distribution over all the spin histories
$p(\vec{\sigma}(0),\ldots,\vec{\sigma}(t))$ has \cut{in principle} the following
simple Markovian form
\begin{equation}
p(\vec{\sigma}(0)\ldots,\vec{\sigma}(t)) = \prod_{s=1}^t \,W[\,\vec{\sigma}(s)\,|\,\vec{h}(s)]\,p(\vec{\sigma}(0))~,
\label{eq:prob_dyn}
\end{equation}
where $W$ is the transition matrix, whose elements are defined by Eq.~(\ref{eq:dynamic}).
The evolution of a a single spin is (trivially) defined by
summing over the histories of all spins except that of spin $i$,
\begin{equation}
p_i(\sigma_i(0),\ldots,\sigma_i(t)) = \sum_{\vec{\sigma}_{\setminus i}(0),\ldots,\vec{\sigma}_{\setminus i}(t)} \, p(\vec{\sigma}(0),\ldots,\vec{\sigma}(t))~,
\label{eq:marginal}
\end{equation}
and similarly for pairwise joint probability of the histories of two spins one omits the corresponding trajectories of $\sigma_i$ and $\sigma_j$ at all times
\begin{equation}
p_{ij}(\sigma_i(0),\ldots\sigma_i(t),\sigma_j(0),\ldots,\sigma_j(t')) = \sum_{\vec{\sigma}_{\setminus i,j}(0),\ldots,\vec{\sigma}_{\setminus i,j}(t)}\, p(\vec{\sigma}(0),\ldots,\vec{\sigma}(t)) \,\,\,\,\, .
\label{eq:correlation}
\end{equation}
%
It can be verified that marginal distribution of spin trajectories [Eq.~(\ref{eq:marginal})]
 can be expressed in terms of the joint probability distribution of neighboring spins
\begin{eqnarray}
p_i(\sigma_i(0),...,\sigma_i(t)) = & p_i(\sigma_i(0))\,\,\displaystyle\sum_{\vec{\sigma}_{\partial i}(0)...\vec{\sigma}_{\partial i}(t-1)}\,
p^{(i)}(\vec{\sigma}_{\partial i}(0),\ldots,\vec{\sigma}_{\partial i}(t)\,\, / \,\, \sigma_i(0),...,\sigma_i(t))&\nonumber\\
&\displaystyle\prod_{s=1}^{t}\,\frac{\exp\left(\beta \sigma_i(s)(\sum_{j\in\partial i}J_{ji}\sigma_j(s-1)+\theta_i(s))\right)}{2\cosh\left(\beta(\sum_{j\in\partial i}J_{ji}\sigma_j(s-1)+\theta_i(s))\right)}\,\,&~.
\label{eq:mar_cav}
\end{eqnarray}
The intuition comes from the concept of cavity in statistical mechanics of disordered systems. The last term on the right hand side is the conditional probability of observing the trajectory $\{\sigma_i(0),\ldots,\sigma_i(t)\}$ when the configuration of neighboring spins is given. The second term, representing the joint probability distribution of neighboring spins, must be evaluated on the cavity graph. Therefore, we denote it by $p^{(i)}(\vec{\sigma}_{\partial i}(0),\ldots,\vec{\sigma}_{\partial i}(t)\,\, / \,\, \sigma_i(0),...,\sigma_i(t))$.
The first term in Eq.~(\ref{eq:mar_cav}) represents the initial conditions for the process.
The site magnetization at time $t$ can be derived from Eq.~(\ref{eq:mar_cav})
 \begin{eqnarray}
m_i(t) =  && p_i(\sigma_i(0))\sum_{\sigma_i(0\rightarrow t)}\,\sigma_i(t)\,\,\displaystyle\sum_{\vec{\sigma}_{\partial i}(0\rightarrow t-1)}\,
p^{(i)}(\vec{\sigma}_{\partial i}(0\rightarrow t)\,\, / \,\, \sigma_i(0\rightarrow t))\nonumber\\
&& \displaystyle\prod_{s=1}^{t}\,\frac{\exp\left(\beta \sigma_i(s)(\sum_{j\in\partial i}J_{ji}\sigma_j(s-1)+\theta_i(s))\right)}{2\cosh\left(\beta(\sum_{j\in\partial i}J_{ji}\sigma_j(s-1)+\theta_i(s))\right)}\,\,~,
\label{eq:mag_cav}
\end{eqnarray}
where we use $(s\rightarrow t)$ to indicate the time path through $\{s,s+1,...,t\}$.
Clearly, computing marginal distribution [Eq.~(\ref{eq:marginal})] and consequently the
magnetization [Eq.~(\ref{eq:mag_cav})] is intractable. It requires a summation of order
 ${\cal O}(2^{T\|c_i\|})$ where $c_i$ is the number of neighboring spins for target site $i$.
We will show that even in the infinite-range spin glass systems, this prohibitive complexity remains.

\section{Infinite-range model}
\label{sec:infinite}
Here, we focus on infinite-range spin glass systems, where the number
of neighbors per spin is large $c_i\propto{\cal O}(N)$, and
 in turn interaction couplings scale inversely to the system size; such that in the thermodynamic
 limit $(N\rightarrow \infty)$ spins become weakly correlated. In the old statistical mechanics
 literature this was the basic idea behind naive mean field theory for the Sherrington-Kirkpatrick (SK) model.
 In order to implement similar assumption in non-equilibrium system, we introduce the
 probability of observing trajectories (time path) for the effective fields [Eq.~(\ref{eq:eff_field})]
\begin{equation}
p_i(h_i(1\rightarrow t)) = \displaystyle\sum_{\vec{\sigma}_{\partial i}(0\rightarrow t-1)} p(\vec{\sigma}_{\partial i}(0\rightarrow t-1)) \,\,\displaystyle\prod_{s=1}^{t}\,\delta(h_i(s)-\sum_{j\in\partial i}J_{ji}\sigma_j(s-1))~.
\label{eq:eff_field_dist}
\end{equation}
Note that introducing this probability will not facilitate our calculations even in the
infinite-range model. In fact, computing $p_i(h_i(1\rightarrow t))$ is as difficult as
computing the joint probability distribution of neighboring spins in Eq.~(\ref{eq:mag_cav}).
This is due to the fact that dynamics of neighboring spins are correlated
 through the target spin $i$. In other words, the target spin $i$ affects its neighboring spins
 and consequently they become correlated.

However, for fully asymmetric networks one can assume that the effective field $h_i(t)$ admits a Gaussian distribution, due to the small correlation between different sites at successive steps~\cite{Mezard2011exact}.
Consequently, the time single-spin magnetization is given by
\begin{equation}
m_i(t) = \int  \,\, Dx
\,\,\tanh\left[\beta\left(\sqrt{\Delta_i(t)}\,\, x+\sum_{j\in\partial i} J_{ji} m_j(t-1)+ \theta_i(t)\right)\right] ~,
\label{eq:Mezard}
\end{equation}
where $Dx = \frac{dx}{\sqrt{2\pi}} e^{-x^2/2}$ is the Gaussian probability density and $\Delta_i(t) = \sum_{j\in i} J_{ji}^2 (1-m_j^2(t-1))$. Note that Eq.~(\ref{eq:Mezard}) is derived under the assumption of vanishing non-diagonal elements of the covariance matrix $(C_{ij}=0 \,\,\, i\neq j)$. Under this assumption the time dependent magnetization admits a Morkovian equation where at each time only information from the previous time step is required.

The results are exact in the thermodynamic limit $(N\rightarrow \infty)$
for any set of couplings and external fields, as it relies only on the central limit theorem.
Equation~\ref{eq:Mezard} should be compared with the naive mean field approximations introduced in~\cite{RoudiHertz2011}
\begin{equation}
m_i(t) = \tanh\left[\beta \left(\sum_{j\in\partial i} J_{ji} m_j(t-1)+ \theta_i(t)\right)\right]~.
\label{eq:nMF}
\end{equation}

The naive mean field equation can be derived from Eq.~(\ref{eq:Mezard}) by an expansion in interaction strength.
In~\cite{aurell2012dynamic} it was shown that such expansion, in the first order, will give rise to Eq.~(\ref{eq:nMF}) and in second order will produce a \emph{TAP-like} equation~\cite{HertzRoudi2011,aurell2012dynamic}.
Here, we will improve on the analysis of the dynamics by taking into account temporal correlations.
The idea is to assume a Gaussian distribution for the effective fields but consider a non-zero covariance matrix. The approach is similar to the one introduced by Amari and Maginu~\cite{amari1988statistical},
where they studied non-equilibrium dynamics of autocorrelation associative memory.
Assuming the Gaussian distribution for the effective field, one can write a general description for the time dependent magnetization
\begin{equation}
m_i(t) =  \int  \,\, Dx
\,\,\tanh\left[\beta\left(\sqrt{V_i(t)}\,\, x+\sum_{j\in\partial i} J_{ji} m_j(t-1)+ \theta_i(t)\right)\right]~,
\label{eq:US}
\end{equation}
where $V_i(t) = \langle h_i^2(t)\rangle - \langle h_i(t)\rangle^2$ is the autocovariance of the effective field at site $i$. In order to perform this integral we first need to compute $V_i$. From the definition of $h_i$ we can substitute the effective field by its explicit expression in terms of spin values at an earlier time step
\begin{equation}
V_i(t) = \sum_{j\in\partial i} J_{ji} J_{ki}\,\,\, \langle\sigma_j(t-1)\sigma_k(t-1) - m_j(t-1) m_k(t-1)\rangle ~.
\end{equation}
If connectivities are exactly evenly distributed or if we assume non-diagonal elements of the covariance matrix to be zero, the above equations reduces to Eq.~(\ref{eq:Mezard}). However, these conditions are not satisfied in general and the non-diagonal elements must be taken into account.
In what follows we calculate $V_i(t)$ in terms of earlier correlation functions taking into account
correlations between spins and effective fields at earlier times. We will show how using
more correlations improves our predictions. By definition, for $V_{i,j}(t,s+1)$ we have
\begin{eqnarray}
V_{i,k}(t,s+1) &=& \langle h_{i}(t) h_k(s+1) \rangle - \langle h_i(t)\rangle\, \langle h_k(s+1)\rangle \nonumber\\
&=& \sum_{j\neq k} J_{jk}\,\,  \left( \langle \sigma_j(s) h_i(t) \rangle - m_j(s) \langle h_i(t)\rangle\,  \right)~.
\end{eqnarray}
The key idea here is to substitute $  \langle h_{j}(s) \sigma_k(t-1) \rangle$ by early time correlations. Without loss of generality we assume that $s\leq t$ and
\begin{equation}
\langle h_{i}(t) \sigma_j(s) \rangle = \sum_{\sigma_j(0\rightarrow s )} \, \int d h_i(t) \,\prod_{s'=1}^s d h_j(s') \,\sigma_j(s)\, h_i(t)\,\, p(\sigma_j(0\rightarrow s), h_j(1\rightarrow s), h_i(t))~,
\end{equation}
where the summation is taken over all possible configurations of $\sigma_j$ at all time steps from $0$ to $s$.
The two random variable $h_i(t)$ and $\sigma_j(s)$ are correlated through $h_j(1\to s)$ in time and the joint probability distribution is denoted by $p(\sigma_j(0\rightarrow s), h_j(1\rightarrow s), h_i(t))$. Now we use Bayes theorem to express the joint probability distribution in a more appropriate form
\begin{equation}
\langle h_{i}(t) \sigma_j(s) \rangle = \sum_{\sigma_j(0\rightarrow s )}  \int d h_i(t)\, \prod_{s'=1}^s d h_j(s')\, \sigma_j(s) \,h_i(t)\,\, p(\sigma_j(0\rightarrow s)| h_j(1\rightarrow s)) \,\, p(h_i^{(j)}, h_j^{(j)})~.
\end{equation}
Here $h_i^{(j)}$ stands for the {\it cavity} effective field acting on spin $i$ when the time path for spin $j$ is fixed.
Since the cavity effective fields $h_i$ and $h_j$ are random variables their joint probability distribution is represented by a multivariate Gaussian
\begin{equation}
 p(h_i^{(j)}, h_j^{(j)}) =  \frac{1}{\sqrt{2\pi V_i V_j(1-x)}} \, \exp\left(-\frac{(h_i^{(j)}-\langle h_i^{(j)}\rangle)^2}{2V_i\left(1-x\right)}-\frac{(h_j^{(j)}-\langle h_j^{(j)}\rangle)^2}{2V_j\left(1-x\right)}+V_{ij}\frac{(h_i-\langle h_i^{(j)}\rangle)(h_j^{(j)}-\langle h_j^{(j)}\rangle)}{V_jV_i\left(1-x\right)}\right) ~,
 \label{eq:joint_prob}
\end{equation}
where $x=V_{ij}^2/(V_iV_j)$ and is small but non-zero in the thermodynamic limit.
In deriving equation~(\ref{eq:joint_prob}) we assume that the cavity covariance function is equal to the normal covariance function {\it i.e.} $V_{ij}^{(j)}=V_{ij}$. This turns out to be correct in the thermodynamic limit $N\to \infty$ for the SK model since the couplings are drawn independently and at random.
The conditional probability $p(\sigma_j(0\rightarrow s)| h_j(1\rightarrow s))$ follows the Glauber dynamics rule defined in Eq.~(\ref{eq:dynamic}).

In order to perform the integral we introduce the explicit expression of the field $h_i(t) = h_i^{(j)}(t)+J_{ji}\sigma_j(t-1)$. We then have
\begin{eqnarray}
 \langle h_{i}(t) \sigma_j(s) \rangle &= & \frac{1}{\sqrt{2\pi V_i V_j(1-x)}} \int d h_i(t) \prod_{s'=1}^sd h_j(s') \,\,  \left(h_i^{(j)}(t)+J_{ji} \sigma_j(t-1)\right) \,\, \sigma_j(s)\, \nonumber\\
 &&  \prod_{s'=1}^s \frac{e^{\beta \sigma_j(s') \left(h_j(s')+\theta_j(s')\right)}}{2 \cosh(\beta\left( h_j(s')+\theta_j(s')\right))}\,\,\,  p(h_i^{(j)}(t), h_j^{(j)}(s))~.
\label{eq:corr_f_asym}
\end{eqnarray}
By performing integration by part in Eq.~(\ref{eq:corr_f_asym}) we get
\begin{equation}
\langle h_{i}(t) \sigma_j(s) \rangle = V_{ij}(t,s)\,\, G_j(s) + m_j(s)\, \langle h_i^{(j)}\rangle(t) + J_{ji} \langle \sigma_j(t-1)\sigma_j(s) \rangle~,
\label{eq:key_asym}
\end{equation}
where $G_i(s)$ is the generalized response function
\begin{equation}
G_i(s) =  \int  \,\, Dx
\,\,\left[ 1 - \tanh^2\left(\beta(\sqrt{V_j(s)}\,\, x+\sum_{k\in\partial j} J_{kj} m_k(s-1)+ \theta_j(s))\right)\right]~.
\end{equation}
Equation~(\ref{eq:key_asym}) is a remarkable result for dynamics of SK model as it relates the correlations between a single spin with the effective fields of its neighbors to the earlier time correlations. Substituting Eq.~(\ref{eq:key_asym}) into Eq.~(\ref{eq:corr_f_asym}) we get
\begin{equation}
V_{i,j}(t,s) = \sum_{k\neq i} J_{ki}\displaystyle \left(\, G_k(t-1)  \, V_{ik}(t-1,s) + J_{ik} C_k(t-1,s-1)\right)~,
\label{eq:it_Vnonmatrix}
\end{equation}
or in matrix notation
\begin{equation}
V(t,s) = J^{\sf T} G(t-1) V(t-1,s) + J^{\sf T} C(t-1,s-1) J~.
\label{eq:it_V}
\end{equation}
where the matrix $C$ on the right hand side of (\ref{eq:it_Vnonmatrix}) and (\ref{eq:it_V}) is the auto-covariance function $C_{ik}(t,s) = \delta_{ik}\left(\langle\sigma_k(t)\sigma_k(s)\rangle - m_k(t)m_k(s)\right)$.
A particular use of Eq.~(\ref{eq:it_V}) is to find an estimate for the auto-covariance function $V_i(t)$ as it is required for evaluating the magnetizations. We therefore express it in terms of covariance function values at different times as
\begin{equation}
V_i(t) =  \sum_{j\neq i,k\neq i} J_{ki} J_{ji}\displaystyle , G_j(t-1) G_k(t-1)  \, V_{jk}(t-1,t-1) + \sum_{k} J^2_{ki} (1-m^2_k(t-1))~.
\end{equation}
Higher order of covariance function can be computed using Eq.~(\ref{eq:key_asym})
\begin{equation}
V(t,t+n) = V(t,t) \, \displaystyle \prod_{s=1}^{n-1}\,\,\, \displaystyle\left[\,\,G(t+n-s)\,J\,\,\right] + \displaystyle \sum_{s=1}^{n-1} J^{\sf T}\,C(t+n-s,t-1)\,J\displaystyle \prod_{s'=1}^{s-1}\left[\,\,G(t+n-s)\,J\,\,\right]~,
\label{eq:cov_V_it}
\end{equation}
where for the special case of $n=1$ we have
\begin{equation}
V(t,t+1) = V(t,t) G(t) J + J^{\sf T} C(t-1,t)\, J~.
\end{equation}	
Note that $V$ is the covariance function for effective fields and is related to the spin covariance $V=J^{\sf T} {\cal C} J$ where ${\cal C}$ is the full covariance matrix. Assuming that  $J^{-1}$ exists we have
\begin{equation}
{\cal C}(t-1,t) = {\cal C}(t-1,t-1) J G(t-1)  + C(t-1,t) \,~,
\label{eq:one_corr}
\end{equation}
where $C(t-1,t)$ is the the autocorrelation ($C_{ij}=0$ for $i\neq j$).
This equation provides us with an approximation for the non diagonal elements of covariance function between pairs of spins in SK model at different times.
\begin{equation}
 {\cal C}_{ij}(t-1,t) = \sum_{k}{\cal C}_{ik}(t-1,t-1) J_{kj} G_j(t-1) \,\,\,\,\,\,\,\,\,\,\, {\rm for \,\,\,i\neq j}~,
 \end{equation}
Diagonal terms of covariance function must be computed through mean field equations. Since the effective fields are randomly distributed (drawn from a Gaussian distribution) the covariance function follows the bivariate Gaussian distribution
\begin{eqnarray}
&& C_{i}(t-1,t) = \frac{1}{\sqrt{det(2\pi V_i)}} \int dh_i(t)\int dh_i(t-1) \tanh(\beta h_i(t) + \beta \theta_i) \tanh(\beta h_i(t-1) + \beta \theta_i) \\
&& \exp\left(-\frac{(h_i(t)-\langle h_i\rangle(t))^2}{2V_i(t)\left(1-x\right)}-\frac{(h_i(t-1)-\langle h_i(t-1)\rangle)^2}{2V_i(t-1)\left(1-x\right)}+V_{i}(t,t-1)\frac{(h_i(t)-\langle h_i(t)\rangle)(h_i(t-1)-\langle h_i(t-1)\rangle)}{V_i(t)V_i(t-1)\left(1-x\right)}\right)~. \nonumber
\label{eq:auto_C}
\end{eqnarray}
We like to point out a major difference between equations~(\ref{eq:one_corr}) and (\ref{eq:auto_C}) for the covariance matrix and the one used in the mean field approximation of fully asymmetric networks~\cite{Mezard2011exact}. For the fully asymmetric mean field approximation, the covariance matrix of spins at two successive time is fully determined by equation~(\ref{eq:one_corr}). However, here the auto-covariance elements must be calculated separately according to equation~(\ref{eq:auto_C}).

The Gaussian assumption for the probability distribution of effective fields is correct only in fully asymmetric networks but it can be used as an approximation in networks with symmetric or partially symmetric connectivities. We can improve the approximation by taking into account the non diagonal elements of covariance function within the Gaussian approximation for the effective fields.
This approximation is similar to~\cite{amari1988statistical} and~\cite{okada1995hierarchy} introduced in the context of statistical neuroscience.


\section{Kinetic inverse Ising problem}
\label{sec:inverse}
Our proposed mean field approximation can be used to solve the non-equilibrium inference problems in densely connected networks. To examine the efficacy of the method we apply it to a network reconstruction inference task - the kinetic inverse Ising problem. The task is to infer coupling interactions $J_{ij}$ and external fields $h_i$ of an Ising model given time dependent magnetizations $m_i(t)$ and covariances $C_{ij}(t,s)$ observed from data. In the setup used we sample the magnetizations and correlations of a system evolving according to the Glauber dynamics~\cite{glauber1963time}.

In the current example, we focus on inferring coupling interactions from $C(t,t-1)$ and $m(t)$. However, the framework used enables one to refine the inferred values by considering a broader range of covariance matrices of various time differences.
We start from Eq.~(\ref{eq:one_corr}); the connectivity matrix appears explicitly on the right hand side and implicitly in the vector $G$. Following~\cite{Mezard2011exact} we obtain the connectivity matrix by inverting equation~(\ref{eq:one_corr}). Introducing $B(t) = J G(t-1)$ we have
\begin{equation}
B(t-1) = {\cal C}^{-1}(t-1,t-1)\left[(1-\delta\right)] {\cal C}(t,t-1)~,
\label{eq:JG}
\end{equation}
where $\left[(1-\delta\right)]{\cal C}(t,t-1)$ contains \emph{only non-diagonal terms} of the covariance matrix ${\cal C}(t,t-1)$. The right hand side is entirely given by data, while the left hand side depends directly on the couplings. In order to compute the connectivities we use an iterative procedure proposed in~\cite{Mezard2011exact}. We express $V(t)$ in terms of known quantities; by definition, $V_i(t)$ is the covariance matrix of the effective fields whereas ${\cal C}$ stands for covariance matrix of the spins. These two matrices are clearly related as
$V(t) = J^{\sf T} \,\, {\cal C}(t-1) \,\, J $.
Multiplying the connectivity matrix $J$ by an identity matrix $GG^{-1}$ from both sides simplifies the equation since $JG$ is given by empirical data [see Eq.~(\ref{eq:JG})]. The final expression for the covariance matrix $V$ is
\begin{equation}
V(t) = G^{-1}(t) \left[\,GJ^{\sf T} \,\, {\cal C}(t-1) \,\, JG\,\right] G^{-1}(t)~.
\end{equation}
The \emph{diagonal elements} of $V(t)$ appear from computing magnetizations and correlations on the right hand side.
At each time step we use the following three set of mean field equations
\begin{eqnarray}
m_i(t) &=&  \int  \,\, Dx
\,\,\tanh\left(\beta(\sqrt{V_i(t)}\,\, x+\sum_{j\in\partial i} J_{ji} m_j(t-1)+ \theta_i(t)\right) \nonumber\\
G_i(t) &=& \int  \,\, Dx
\,\,\left[ 1 - \tanh^2\left(\beta\left(\sqrt{V_i(t)}\,\, x+\sum_{j\in\partial i} J_{ji} m_j(t-1)+ \theta_i(t)\right)\right)\right] \nonumber \\
V_i(t) &=& \frac{{\cal D}_i}{G_i^2(t)}~, \nonumber
\label{eq:rec_GMF}
\end{eqnarray}
where ${\cal D}_i = \sum_{j,k} B_{ij}(t) C_{jk}(t-1) B_{ki}(t) $ is calculated directly from the data as mentioned before.
We start from an initial value for $\Delta_i(t)\equiv \sqrt{V_i(t)}$. Given $m_i(t)$, from the empirical data, we solve the first equation to find $\sum_{j\in\partial i} J_{ji} m_j(t-1)+ \theta_i(t)$. We then substitute $\sum_{j\in\partial i} J_{ji} m_j(t-1)+ \theta_i(t)$ into the second equation to compute $G_i(t)$. The third equation updates $\Delta_i(t)$. Once the iterative process has converged we use $G_i(t)$ and $V_i(t)$ to extract connectivities and external fields values from Eq.~(\ref{eq:JG}).

Note that the only difference between our generalized mean field theory and the one proposed in~\cite{Mezard2011exact} is the third equation. In our method we use Eq.~(\ref{eq:JG}) to compute the data, whereas for the mean field method of fully asymmetric networks~\cite{Mezard2011exact} $V_i(t)$ follows a simpler equation $V_i(t)= \sum_{j\in i} J_{ji}^2 (1-m_j^2(t-1))$. For the naive mean field approximation~\cite{HertzRoudi2011} the reconstructed connectivities are given by
\begin{equation}
J = A \,\,C(t,t-1)\,\, C^{-1}(t-1,t-1)~,
\label{eq:rec_nMF}
\end{equation}
where $A= \sum_{j\in i} J_{ji}^2 (1-m_j^2(t-1))$.


\section{Results}
\label{sec:results}

We investigate the performance of generalized mean field approximation both for calculating macroscopic quantities (forward problem) and inferring coupling values from data (inverse problem). For the former, given connectivities and external fields we compute time-dependent magnetization, auto-covariance and covariance matrix for fully connected asymmetric networks evolving via parallel updates.
We compare our results to the mean field approximation introduced for the fully asymmetric networks~\cite{Mezard2011exact} and the naive mean field approximation~\cite{RoudiHertz2011}.
We evaluate results obtained by the different methods against numerical simulations based on the Glauber dynamics defined in Sec.~\ref{sec:model}.

For the kinetic inverse Ising problem we sample a network instance at random, and generate
correlations and magnetizations by heavily sampled Glauber dynamics. We then infer the connectivities values based on the different mean field methods described above.

\subsection{Forward problem}
In this case network connectivities are known and time dependent magnetizations are computed according to equations~(\ref{eq:Mezard}),(\ref{eq:nMF}) and (\ref{eq:US}) for the mean field, naive mean field and generalized mean field methods, respectively.
To compute the time dependent covariance matrix we use Eq.~(\ref{eq:one_corr}) in the generalized mean field method. Note that Eq.~(\ref{eq:one_corr}) provides results only for non diagonal terms. The auto-covariance functions can be estimated by Eq.~(\ref{eq:auto_C}). Note also that we can not compute the covariance function by the naive mean field method and mean field method for the fully asymmetric networks. Higher order of covariance functions can be also computed by Eq.~(\ref{eq:cov_V_it}).

\begin{figure}[htb]
\vspace{0.2cm}
\includegraphics[width=0.43\columnwidth]{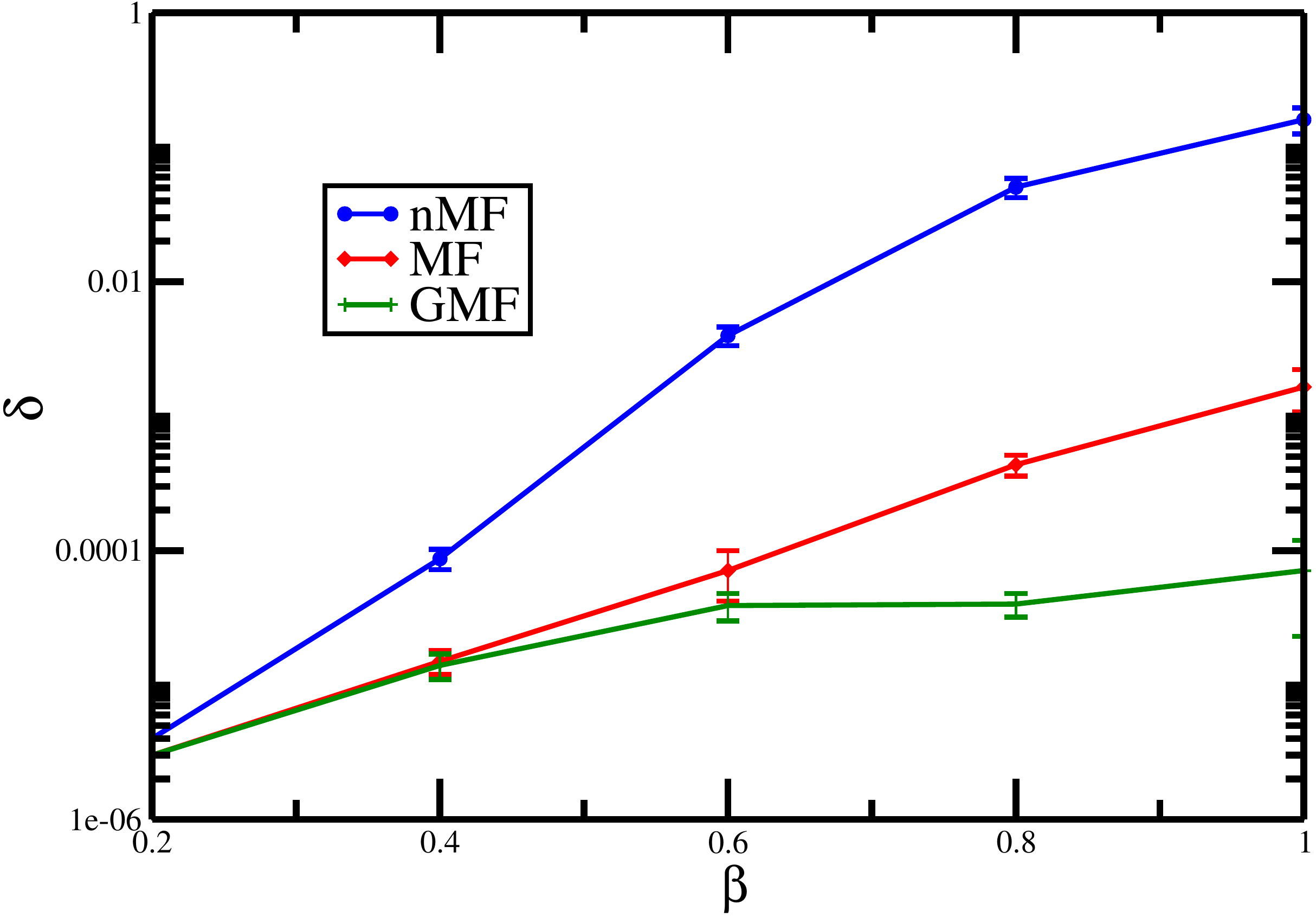}
\hspace{10mm}
\includegraphics[width=0.43\columnwidth]{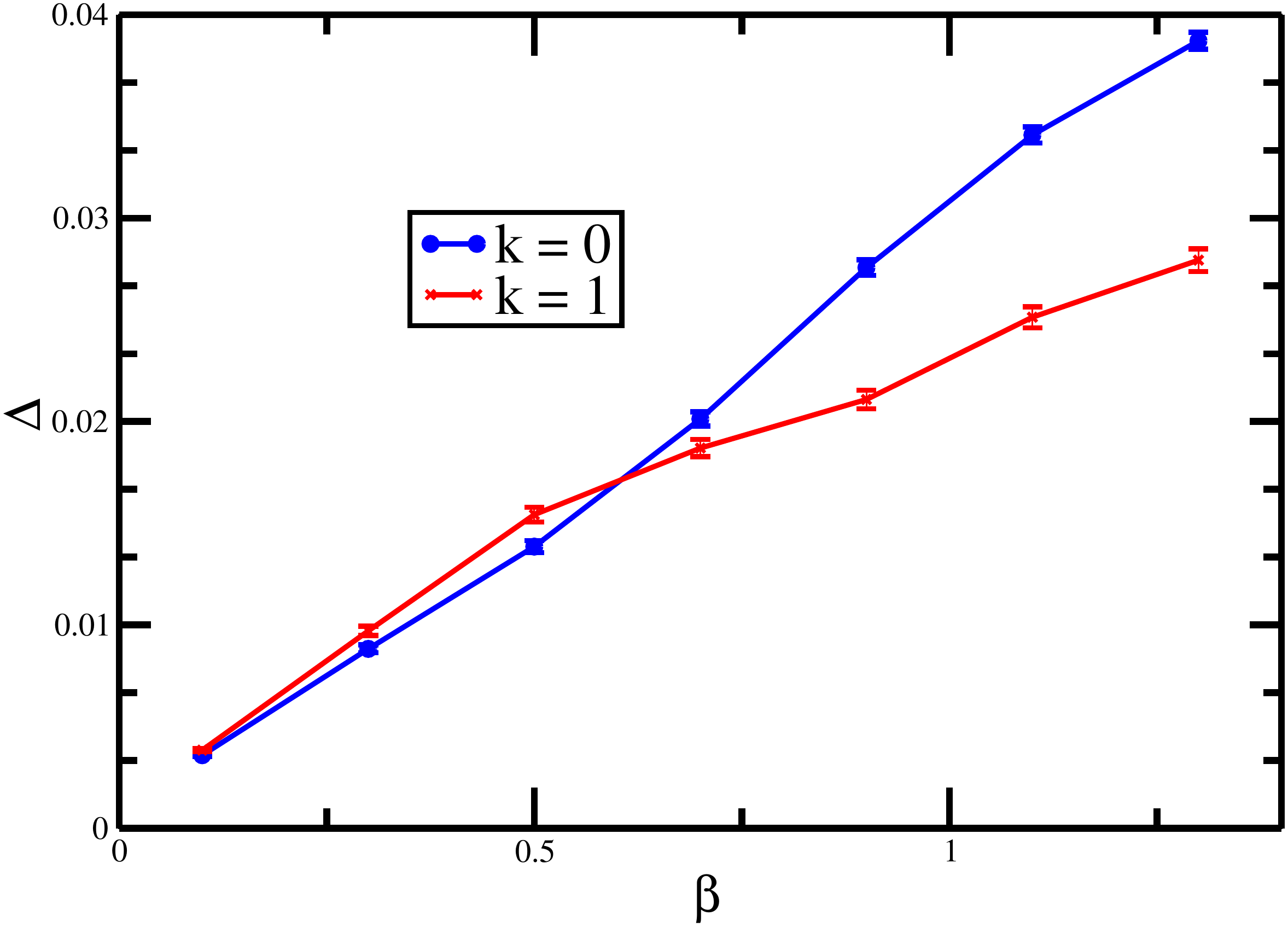}
\caption{Left panel: Squared deviation of  spin averages $\bar{\delta}$ evaluated by different methods for a fully symmetric network $(k=0)$. The results are averaged over 10 time steps.
Right panel: Squared deviation of covariance function computed by generalized mean field (GMF) method. Results are shown for both symmetric and asymmetric networks.
In all simulations, $N=100$, $T=10$ and the sample size for the numerical simulations is $10^6$. Results are averaged over 10 network instances, both mean values and standard deviations are shown (color online).}
\label{fig:mse_mc}
\end{figure}

We examine the performance of different methods by computing the mean squared error defined as
\begin{eqnarray}
\delta(t) &=&  \frac{1}{N}\sum_{i=1}^N\, (m_i^{\rm GD}(t)-m_i^{{\rm estimate}}(t))^2 \nonumber ~,\\
\Delta(t) &=&  \frac{1}{N (N-1)}\sum_{i,j}^N\, (C_{ij}^{\rm GD}(t,t-1)-C_{ij}^{{\rm GMF}}(t,t-1))^2~,
\end{eqnarray}
where $\delta(t)$ is the mean squared error of time-dependent magnetizations obtained by the different methods and $\Delta(t)$ the corresponding mean squared error for the covariance matrices obtained. Here $m_i^{\rm GD}$ and $C_{ij}^{\rm GD}$ stand for magnetization and covariance obtained by numerical simulations of systems evolving via Glauber dynamics, respectively. Note that for $\Delta(t)$ only generalized mean field approximation is considered as covariance matrices cannot be calculated by the other two methods.
Fig.~\ref{fig:mse_mc} shows results for mean squared errors $\delta$ and $\Delta$ averaged over $T$ time steps (here $T=10$).
Left panel compares averaged mean squared errors for magnetization $\bar{\delta} = 1/T \sum_{t=1}^T \delta(t)$ computed by the different methods for fully symmetric networks.
Clearly, generalized mean field approximation performs better for the various $\beta$ values.
The mean field approximation [Eq.~(\ref{eq:Mezard})] estimates correctly the magnetizations for small $\beta$ values but the error grows for lower temperatures.
The right panel presents mean squared error results for the averaged covariance function at two successive times $\bar{\Delta} = 1/T \sum_{t=1}^T\Delta(t)$ computed by the generalized mean field approximation for both fully asymmetric and fully symmetric networks. For small $\beta$ values, the covariance matrices predicted by our method are similar to those obtained by
 numerical simulations. As $\beta$ increases it fails in predicting time dependent covariance for fully symmetric networks.

\begin{figure}[htb]
\vspace{0.2cm}
\includegraphics[width=0.43\columnwidth]{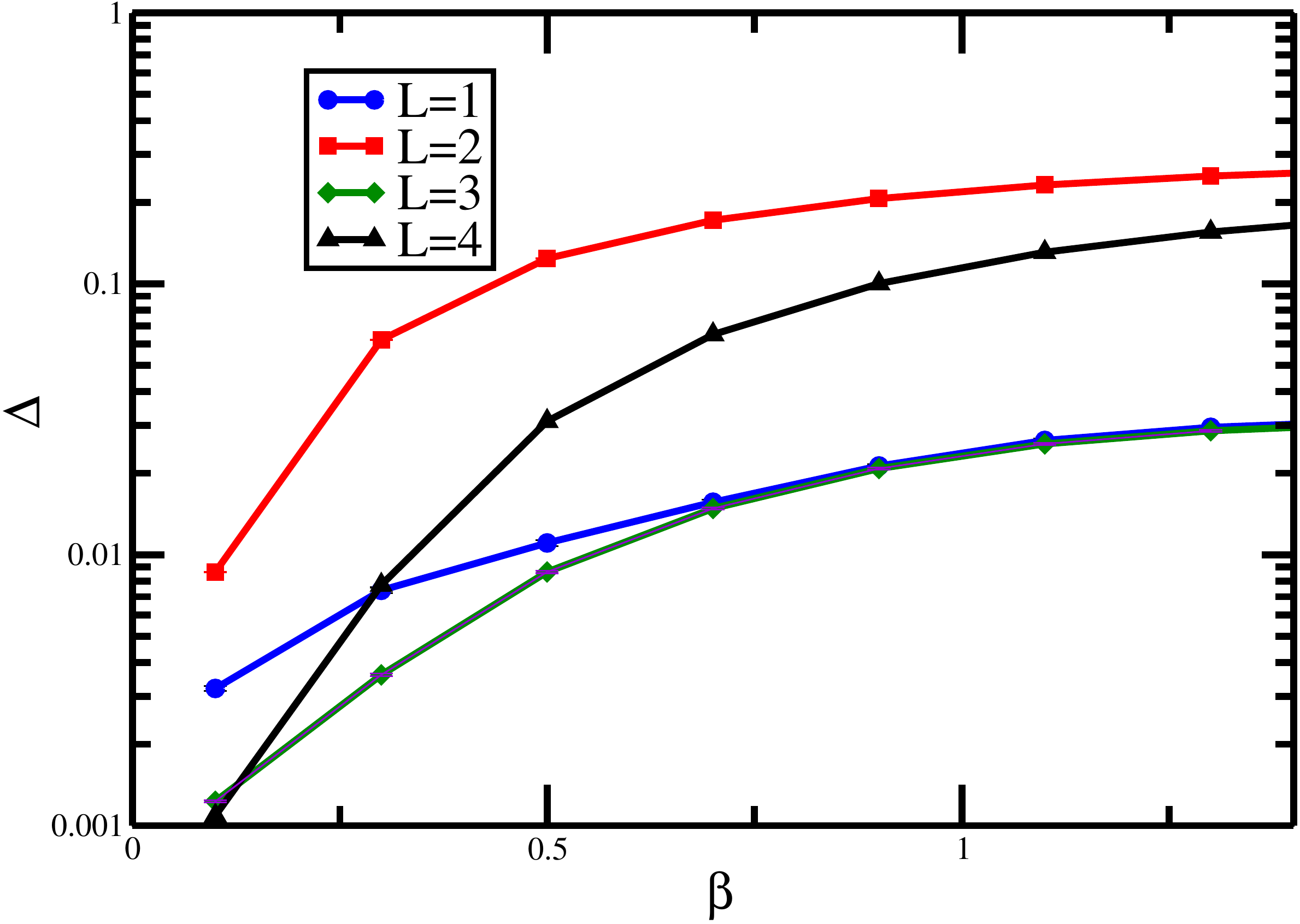}
\hspace{10mm}
\includegraphics[width=0.43\columnwidth]{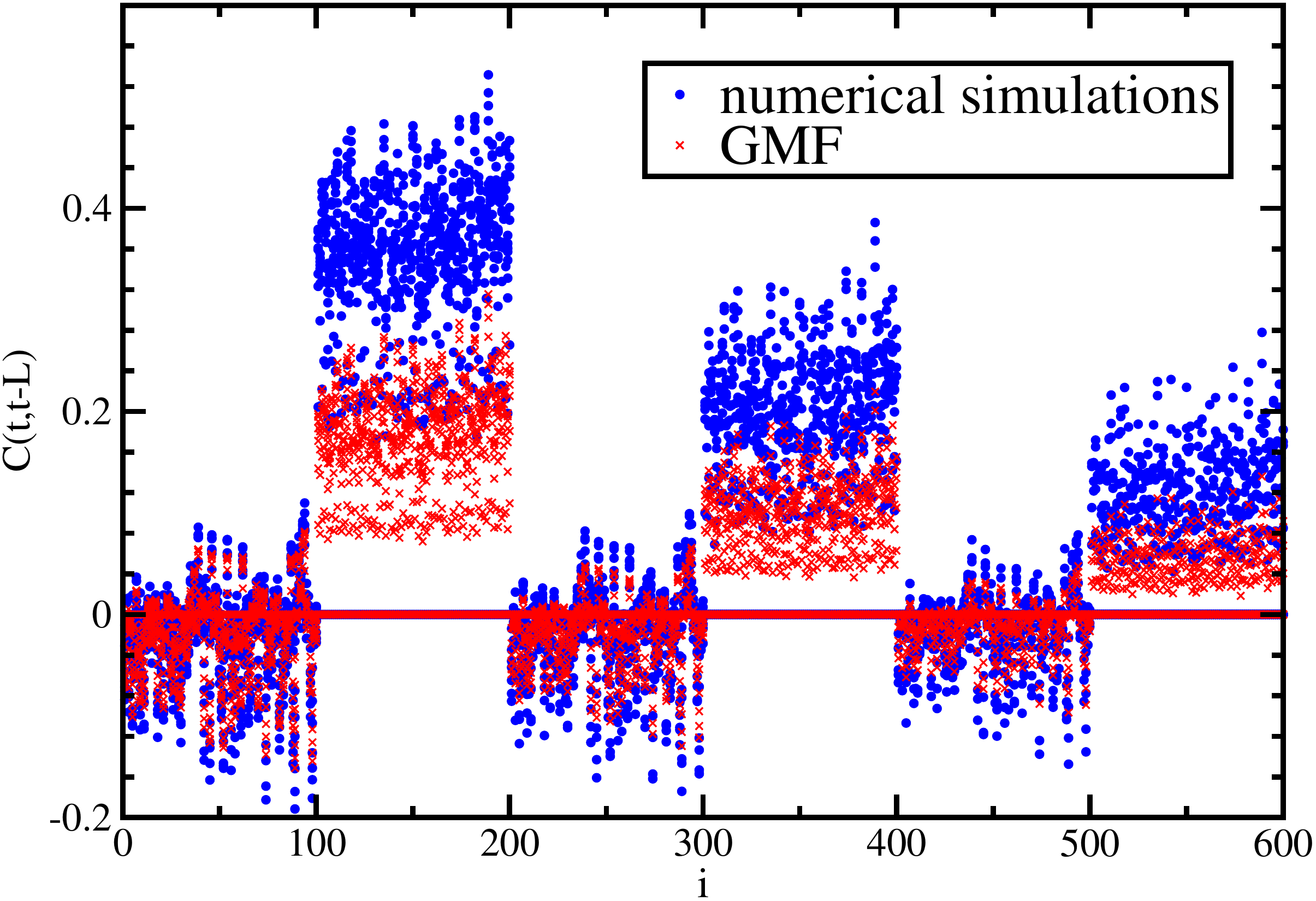}
\caption{Left panel: Squared deviation of spins auto-covariance as a function of $\beta$.
Results are obtained by the generalized mean field method for symmetric networks and are averaged over 10 network samples and 10 time steps. Error bars are smaller than the symbols size.
System size is $100$ and external fields are set to zero.
Right panel: Point-wise auto-covariance function $C_i(t,t-L)$ for different values of $L$ in one single network sample of symmetric networks. The system size is $N=100$ and $\beta = 0.7$. Calculated values are shown in red while values from numerical simulations are shown in blue (color online).}
\label{fig:mss_auc2}
\end{figure}

Figure~\ref{fig:mss_auc2} shows the performance of the generalized mean field method in calculating the auto-covariance function $C_i(t,t-L)$ for various values of $L$ in fully symmetric networks. Due to the parallel nature of dynamics, even values of $L$ give rise to higher auto-covariance values and consequently higher errors.

\begin{figure}[htb]
\hspace{10mm}
\includegraphics[width=0.7\columnwidth]{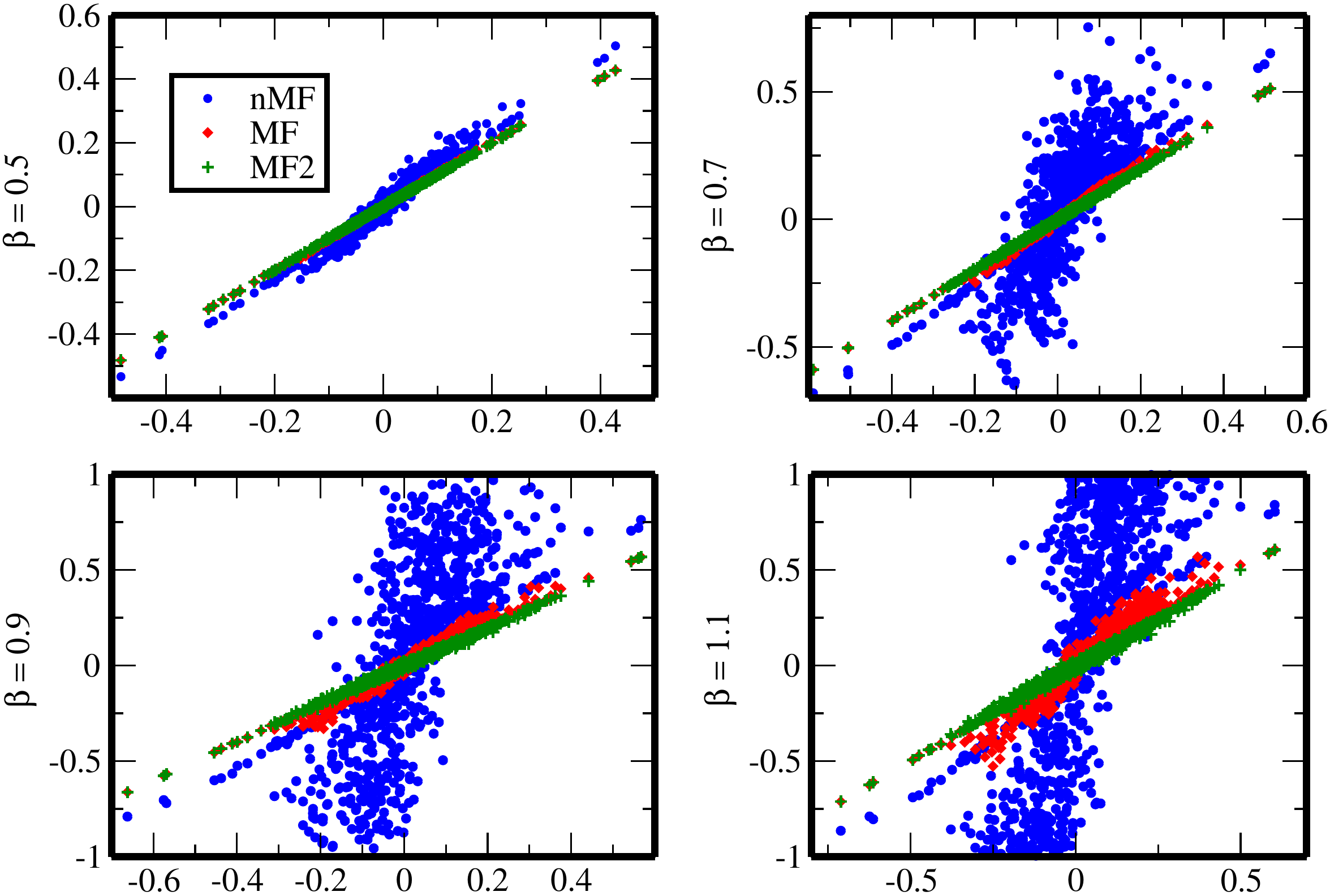}
\caption{Scatter plot of local magnetizations for fully symmetric networks($N=100$) obtained by various methods, plotted against values obtained from simulation ($x$-axis).
Green points represent results obtained by the generalized mean field method, blue points by naive mean field and red points by the mean field method (marked as MF2, color online).
Results are for $\beta = 0.5, 0.7, 0.9, 1.1$ respectively.}
\label{fig:scatter_mag}
\end{figure}

The left panel in Fig.\ref{fig:mss_auc2} shows the mean squared error results of the calculated auto-covariance function. As $\beta$ increase the prediction becomes less accurate. The right hand side panel shows the point-wise auto-covariance function for different $L$ values. Since spins are correlated in a parallel manner, the auto-covariance between two successive time steps is small. However for even $L$ values we observe non-zero auto-covariance values. The results are compared to the numerical simulations (red points). We observe that the generalized mean field method almost always over estimates the auto-covariance function values.
To highlight a qualitative comparison between different methods we show a scatter plots for the local magnetizations obtained by the different methods versus those obtained from numerical simulations in Fig.~\ref{fig:scatter_mag}. The results are for different values of $\beta =\{0.5,0.7, 0.9, 1.1\}$. For small $\beta$ values all methods predict the local magnetizations accurately. As we increase the value of $\beta$, naive mean field starts failing in its prediction. In all cases, generalized mean field theory outperforms the other two methods.
\subsection{Inverse problem}
We now examine the performance of the different methods for the inverse Ising inference problem.
We start with a fully connected spin glass systems with zero external fields and randomly sampled interaction values from Gaussian distribution of zero mean and unit variance;
and then generate time dependent magnetization and covariance functions by applying Glauber dynamics. We examine both cases of symmetric and asymmetric couplings.
The connectivities can be inferred by using Eqs.~(\ref{eq:rec_nMF}) and (\ref{eq:rec_GMF}).
Similar to the previous section we define mean squared error measure for the estimated error in the inferred connectivities
\begin{equation}
{\cal I} =  \frac{1}{N(N-1)}\sum_{ij} \left( J_{ij} - J_{ij}^{\rm rec}\right)^2
\end{equation}
where $J_{ij}^{\rm rec}$ are the estimated (reconstructed) connectivities inferred by the corresponding mean field method and $J_{ij}$ are the original couplings.

\begin{figure}[htb]
\vspace{0.2cm}
\includegraphics[width=0.43\columnwidth]{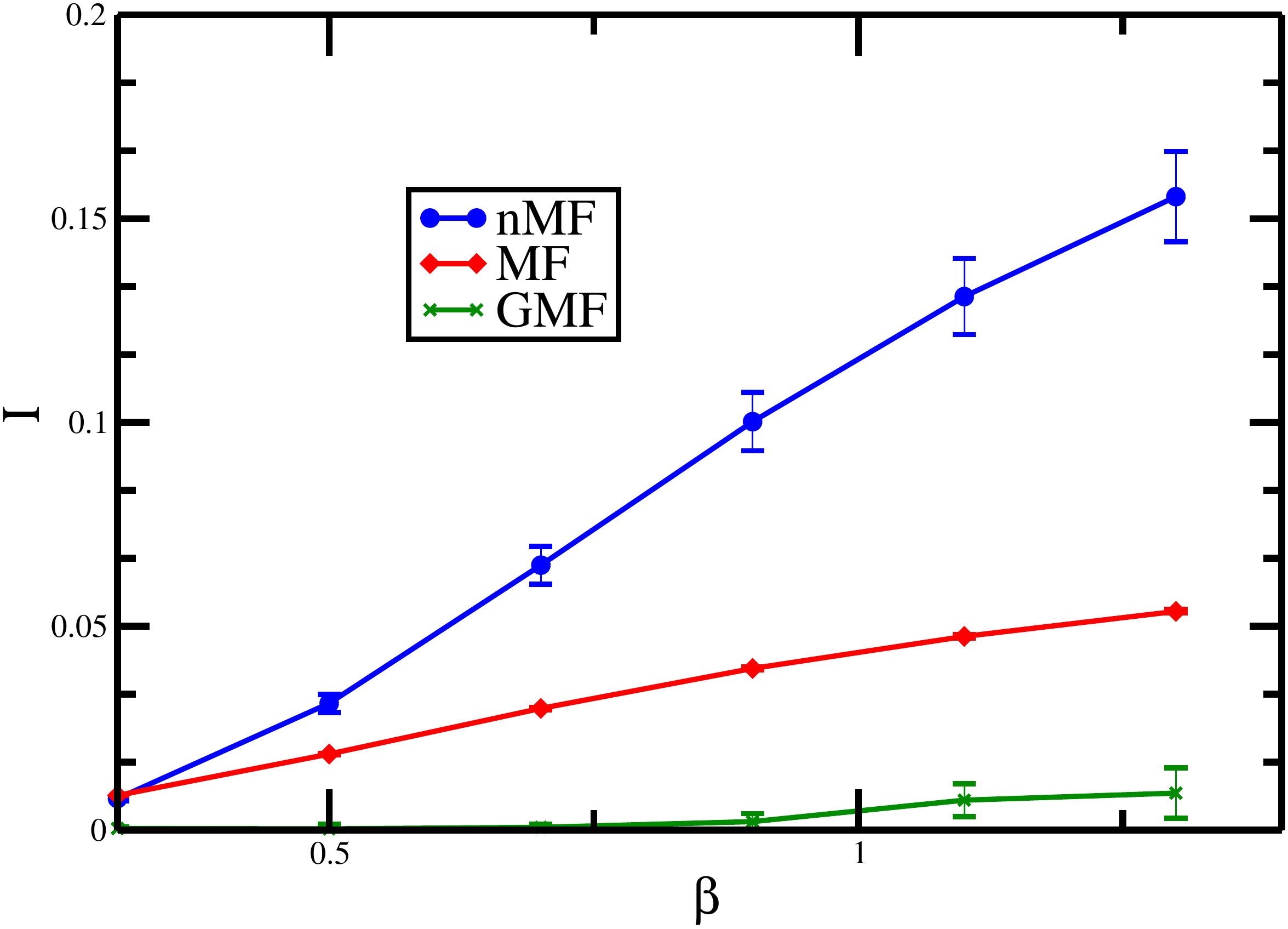}
\hspace{10mm}
\includegraphics[width=0.43\columnwidth]{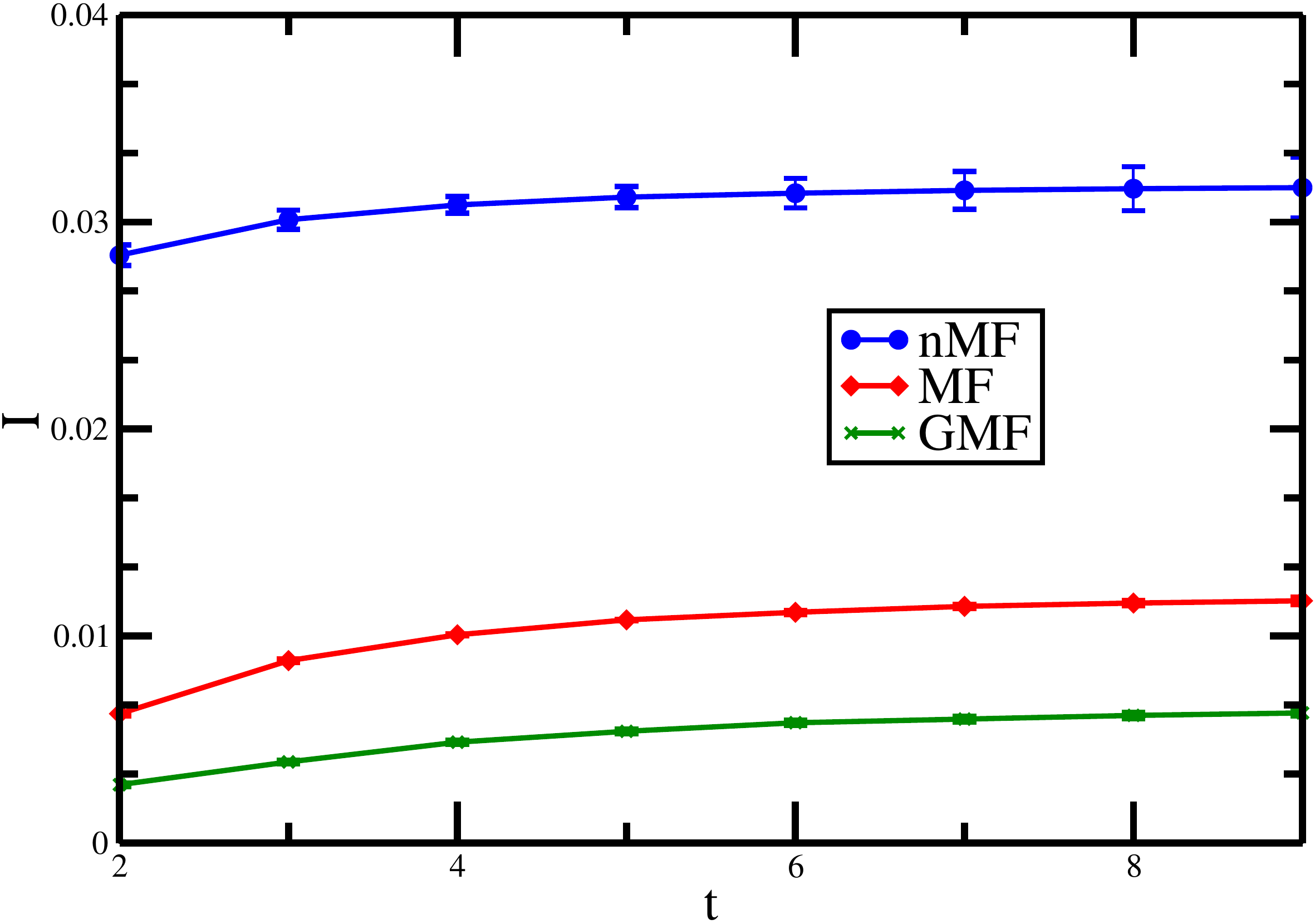}
\caption{Mean squared error for the reconstructed connectivities - ${\cal I}$ in symmetric networks with $N=100$ averaged over 10 network instances.
Left panel shows the performance of different methods for various $\beta$ values averaged over 10 consecutive time steps. Right panel is the mean squared error for $\beta =0.7$ in the different time steps. For results obtained by mean field method and generalized mean field method error bars are smaller than the symbols size (color online).
 }
\label{fig:mss_J}
\end{figure}

\begin{figure}[htb]
\hspace{10mm}
\includegraphics[width=0.43\columnwidth]{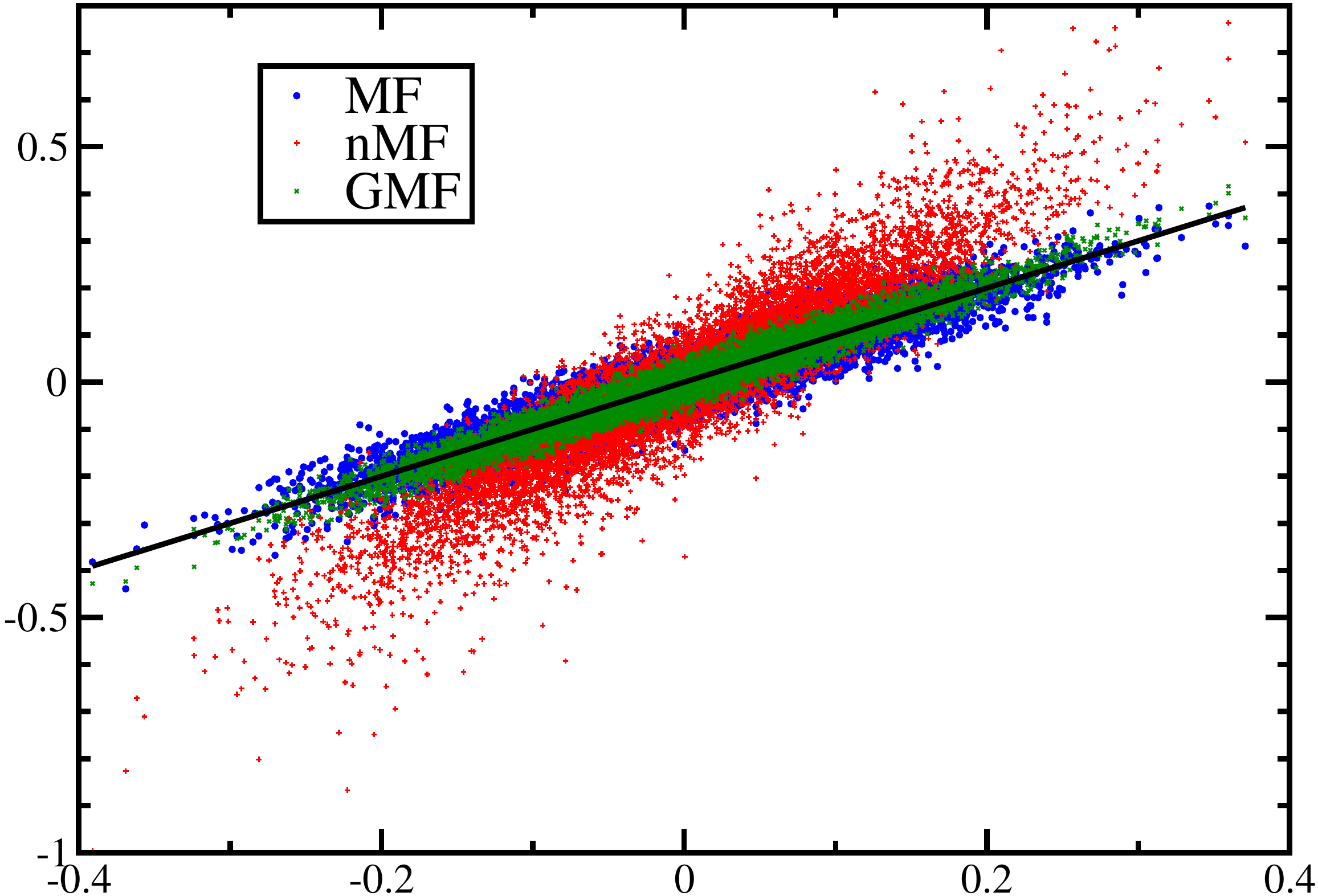}
\caption{Scatter plot of reconstructed connectivities in fully symmetric networks with $N=100$ and $\beta = 0.7$.
Green points represent reconstructed connectivities obtained by the generalized mean field method, blue points by mean field method of fully asymmetric networks and red points by the naive mean field method (color online).
}
\label{fig:scatter}
\end{figure}

The results, presented in Fig.\ref{fig:mss_J}, show that the generalized mean field approximation outperforms the other two methods for different $\beta$ values in symmetric networks. The results are averaged 10 time steps and over 10 networks samples.
Indeed, it provides accurate results (compared to other methods) even for considerably high $\beta$ values.
To observe the comparison in a more qualitative manner,
we show the scatter plot of reconstructed connectivities versus the true connectivities in Fig.~\ref{fig:scatter}.
The generalized mean field method infers perfectly the connectivity couplings for
fully symmetric network even at high temperature, especially compared to mean field and naive mean field results.

\section{Conclusion}
\label{sec:conclusion}
In this paper, we studied the dynamics of spin glass systems and devised an inference method for the inverse Ising model. We focused on the Sherrington-Kirkpatrick model and generalized the mean field approximation, originally developed for the fully asymmetric networks, by taking into account information from the earlier time steps.

A key feature of fully asymmetric networks is to have negligible correlations between non-identical spins at different time steps.
In our proposed method, the correlation between microscopic states at two time steps have been incorporated into the dynamics.
The modification to the mean field approximation shows a great improvement in predicting local properties of the system in both forward and inverseproblems; the former aims at predicting macroscopic properties dynamically, while the latter aims at inferring microscopic values from data in specific instances. Another advantage of our proposed method is that it enables one to compute time dependent correlations at different time steps, a quantity that cannot be studied by other mean field approximations. The performance of our method is evaluated by extensive numerical simulations.

Our numerical experiments show that for the fully asymmetric networks the improvement due to the generalization, over the mean field approximation, is modest as the correlation among the spins at different time steps is small. However, for symmetric networks our method shows significant improvement over other methods and an excellent agreement with data obtained numerically using Glauber dynamics, even at high temperature regime.

This technique can be used to include correlations at any depth, time differences, in a systematic approach to improve the prediction of dynamically evolving macroscopic quantities and inference of microscopic model parameters from data; this is expected as correlations at different times do carry valuable information for both tasks. Extending the calculation to devise improved approximations for symmetric densely connected spin systems is underway as well as the generalization of the method for sparse systems. Both research directions are highly promising and would impact on both theoretical research an practical applications.

This work is supported by the EU project STAMINA (FP7-265496).
\bibliography{spinglas_rev}

\end{document}